\documentclass[12pt,aps,prb,preprint,showpacs,showkeys,groupedaddress]{revtex4}
\usepackage{amssymb,graphicx,psfrag,url}
\begin{document}
\title{A Pedagogical ``Toy'' Bistable Climate Model}
\author{J. I. Katz}
\email[]{katz@wuphys.wustl.edu}
\affiliation{Department of Physics and McDonnell Center for the
Space Sciences \\ Washington University, St. Louis, Mo. 63130}
\date{\today}
\begin{abstract}
A ``toy'' model, simple and elementary enough for an undergraduate class,
of the temperature dependence of the greenhouse (mid-IR) absorption by
atmospheric water vapor implies a bistable climate system.  The stable
states are glaciation and warm interglacials, while intermediate states
are unstable.  This is in qualitative accord with the paleoclimatic data.
The present climate may be unstable, with or without anthropogenic
interventions such as CO$_2$ emission, unless there is additional
stabilizing feedback such as ``geoengineering''.
\end{abstract}
\pacs{92.70.-j, 92.70.Aa, 92.70.Gt}
\keywords{climate, opacity, water vapor, radiative feedback}
\maketitle
\section{Introduction}
Models of climate change, a subject that concerns time scales from decades
to billions of years, range from the simple and qualitative to atmospheric
and oceanic general circulation models running on the fastest computers.
Despite several decades of effort by many scientists, the causes of such
important phenomena as the alternation of glaciation and interglacials, and
the ice ages themselves, remain controversial.  In this pedagogical paper I
describe an elementary and qualitative model of the consequences of one
important effect, the temperature dependence of the opacity of atmospheric
water vapor, the most important greenhouse gas\cite{KT97}.  This is a
consequence of the temperature dependence of the water vapor pressure and
the physics of pressure broadening of saturated Lorentzian spectral line
profiles, and lies on the border between physics and geophysics.
\section{Paleoclimatic Data}
\begin{figure}
\begin{center}
\includegraphics[width=5in]{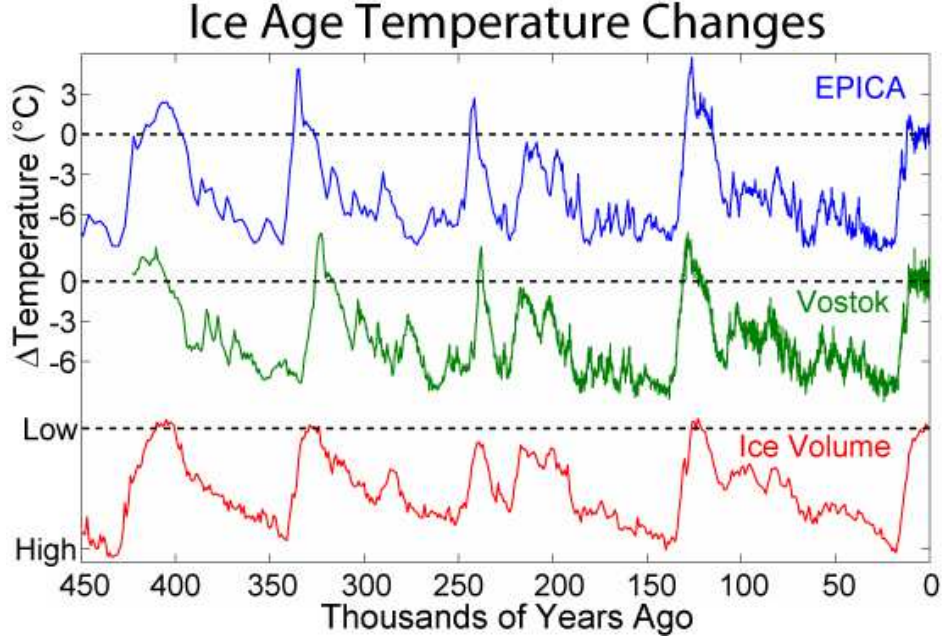}
\caption{\label{data} Paleoclimatic data showing ice ages and
interglacials\cite{W09,E04,P99,LR05}.}
\end{center}
\end{figure}

The paleoclimatic data are summarized in Figure \ref{data}.  Alternation of
ice ages and interglacials is evident.  Intermediate states, such as the
present climate, are not stable or steady, but show continual variation.
This is also evident in the historical record of the Late Medieval Climatic
Maximum ({\it c.\/} 800--1300), followed by the Little Ice Age ({\it c.\/}
1300--1800), followed in turn by a warming trend.
\section{Water Vapor Opacity}
Although water vapor is the greatest source of atmospheric opacity in the
mid-infrared (8--14 $\mu$) band in which the Earth's thermal emission peaks,
it is not even mentioned in a classic monograph\cite{S02} on climatological
dynamics.  The HITRAN database\cite{R09} is the comprehensive source of
spectral information.  Measurements of atmospheric opacity, such as those
shown in Fig.~3 of Hiriart and Salas\cite{HS07}, illustrate its sensitivity
to even small amounts of water vapor. 

Tropospheric water vapor is not usually considered in the context of
anthropogenic greenhouse gas (GHG) emissions because it equilibrates with
surface water very rapidly, with a characteristic time of a few days or
less, and any anthropogenic contribution has no long term effect.
Tropospheric water vapor opacity is controlled by evaporation and
precipitation to a level determined by the surface temperature.
\section{Limit States}
This sensitivity of opacity to water vapor pressure, combined with the
sensitivity of equilibrium vapor pressures to temperature, qualitatively
suggests a bistable system: There may be a cold state in which the water
precipitates and the atmosphere is transparent, with little water vapor 
absorption, and a warm state in which the atmosphere is warm, humid and
opaque and in which absorbed solar energy is carried through the troposphere
by convection (resembling present tropical or temperate summer conditions).

These limit states are stable: in the cold state perturbations to the
temperature have little effect on the infrared opacity because it remains
very low, while in the warm state the opacity has little effect on the heat
transfer.  In either case there is little feedback to the heat balance from
changes in the tropospheric water vapor content.  At intermediate
temperatures increases in temperature produce a positive feedback by
increasing the infrared opacity (and {\it vice versa\/} for decreases), so
that perturbations grow exponentially until one of the stable limit states
is approached.
\section{Model}
Quantitative calculation is precluded by the complexity of the climate 
system.  These complexities includ processes that are difficult to calculate
but understood in principle (such as opacity resulting from a dense comb of
infrared vibration-rotation lines) and those not understood (such as cloud
formation and greenhouse gas exchange with the ocean, soil and biosphere).
Even the most elaborate and sophisticated general circulation models (GCM)
contain many uncertain parameters that can only be calibrated at a single
measured point: today's climate.  It is not surprising that different GCM
disagree in the magnitude of the effect of GHG emission, and it is not
evident how to resolve these disagreements. 

Disagreements among the best state-of-the-art GCM suggest considering the
opposite approach, a model na\"\i ve almost to the point of triviality. If
its workings are transparent, it may provide useful insight, even though it
cannot make quantitative predictions.

Consider a surface heat reservoir warmed by sunlight with a mean (diurnally
and annually averaged) intensity $P_\odot$ and cooled by black body
radiative emission with a (Planck weighted) fraction $f_b(T)$ of the
infrared spectrum blocked by lines of Lorentzian lineshape\cite{G64}
(Doppler broadening of molecular lines is negligible at atmospheric
temperatures).  The optical depth at the centers of strong lines  $\tau_0
\gg 1$, and assume an underlying heat capacity (per unit area) $C_p$.  The 
temperature $T$ of the thermal reservoir changes at a rate
\begin{equation}
\label{temp}
C_p {d T \over dt} = P_\odot - \sigma_{SB} T^4 \left(1 - f_b(T)\right),
\end{equation}
where $\sigma_{SB}$ is the Stefan-Boltzmann constant and we approximate the
line profile as completely opaque within a blocked fraction $f_b(T)$ of
the spectrum, and completely transparent outside it.

We suppose an initial equilibrium temperature $T_0$ defined by the heat
balance condition:
\begin{equation}
0 = P_\odot - \sigma_{SB} T_0^4 \left(1 - f_b(T_0)\right).
\end{equation}
Small deviations of temperature are described by
\begin{equation}
T = T_0 + \delta T.
\end{equation}
\section{Instability}
Expanding Eq.~\ref{temp} to first order in $\delta T$:
\begin{equation}
{1 \over \nu_0 T_0}{d \delta T \over dt} \approx \left[{d f_b(T) \over dT}
- {4 \over T_0}\left(1 - f_b(T_0)\right)\right]\delta T,
\end{equation}
where
\begin{equation}
\nu_0 \equiv {\sigma_{SB} T_0^3 \over C_p}.
\end{equation}
$\delta T$ grows or decays with an exponentiation rate $\nu$:
\begin{equation}
\label{growth}
{\nu \over \nu_0} = T_0 {d f_b(T) \over dT} - 4 \left(1 - f_b(T_0)\right).
\end{equation} 

To estimate the derivative in Eq.~\ref{growth} we use an approximate 
equation \cite{R65} for the saturation water vapor pressure over a surface
of temperature $T$:
\begin{equation}
p_v(T) \propto T^{3/2} \exp{(-\Delta H / k_B T)},
\end{equation}
where $\Delta H = 2260$ J/g is the enthalpy of evaporation of water and
$k_B$ is Boltzmann's constant.  For a Lorentzian line profile with
$\tau_0 \gg 1$, the width and $f_b$ at a specified optical depth
${\cal O}(1)$ 
\begin{equation}
\label{width}
f_b(T) \approx f_b(T_0) \left({\tau_0(T) \over \tau_0(T_0)}\right)^{1/2}
\approx f_b(T_0) \left({p_v(T) \over p_v(T_0)}\right)^{1/2}.
\end{equation}

Then
\begin{equation}
\left.{d f_b(T) \over dT}\right|_{T_0} = {1 \over T_0} f_b(T_0)
\left({\Delta H \over 2 k_B T_0} + {3 \over 4}\right)
\end{equation}
and
\begin{equation}
{\nu \over \nu_0} = f_b(T_0) \left[{\Delta H \over 2 k_B T_0} + {3 \over 4}
- 4 \left(1 - f_b(T_0)\right)\right],
\end{equation}
where we have made the tacit assumptions $1 - f_b(T_0) > \tau_0^{-1/2}$,
which amount to assuming that there is enough unblocked spectrum for the
line to broaden according to Eq.~\ref{width}.  The appropriate value of
$\tau_0$ in this condition is an average over all lines that are optically
thick at their cores.  This is likely to be dominated by the very numerous
lines for which $\tau_0$ exceeds unity by a factor of a few, but not by
orders of magnitude, so these results may be valid for $f_b \lesssim 0.7$.
The model breaks down as $f_b \to 1$ because then there is little unblocked
spectrum into which the lines can broaden.

Evaluating at a sea surface $T_0 = 290^{\,\circ}$K,
\begin{equation}
{\nu \over \nu_0} = 13.2 f_b(T_0) - 4.
\end{equation}
A quantitative calculation of $f_b$ is not possible within this qualitative
model, but spectral data\cite{HS07} suggest values of several tenths.  Then
deviations from equilibrium grow exponentially for
\begin{equation}
0.3 < f_b \lesssim 0.7.
\end{equation}

The mean ocean depth, averaged over the entire Earth (including zero depth
for land areas) is about 2.65 km.  However, the flow of heat to the deep
ocean, while apparently reflecting the warming of the last century
\cite{D09}, is expected to be confined to downwelling regions and to be
comparatively slow.  For the purpose of an order of magnitude estimate we
suppose a nominal mean depth of 1 km contributes to $C_p$ and find
\begin{equation}
\nu_0 = 1.04 \times 10^{-2} / {\rm y}.
\end{equation}
\section{Conclusion}
If $0.3 < f_b \lesssim 0.7$, as suggested by the empirical water vapor
spectrum\cite{HS07}, then instability would be expected for any state
between an ice age (when $p_v$ is low enough that the assumption $\tau_0
\gg 1$ is not satisfied) and the warmest interglacials (when the assumption
$1 - f_b(T_0) > \tau_0^{-1/2}$ is not satisfied as nearly the entire thermal
infrared spectrum is blanketed with water lines and $f_b \to 1$, or energy
is carried by convection, ignored in this na\"\i ve model).  Hence the 
climate would be stable only in these two limiting states.

The paleoclimatic data, as shown in Figure \ref{data}, are qualitatively
consistent with this model.  The historic record of the Medieval Climatic
Maximum, the Little Ice Age, and fluctuations on similar time scales of
${\cal O}(100)$ years at other epochs, is also qualitatively consistent with
the estimated $\nu_0$.
\section{Discussion}
Climatologists have struggled with the question of what upsets the stable
ice age and interglacial states since the discovery of the ice ages, and
this na\"\i ve model cannot contribute to its answer.  Even the most
complete and sophisticated general circulation models struggle with this
problem, and its resolution remains controversial.  However, even the
conclusion that intermediate states may be intrinsically unstable has
significant implications.  One implication is that there is no ``normal''
climate, nor can the Earth be expected to return to such a state even
without anthropogenic interventions such as the emission of GHG.  A second
implication is that if some climate state is determined to be optimal for
humanity, maintaining it would require continual active intervention
(geoengineering).
\begin{acknowledgments}
I thank R.~W.~Cohen, M.~Gregg, W.~Happer, R.~Muller, W.~Munk and C.~Wunsch
for useful discussions. 
\end{acknowledgments}
\bibliography{bistableclimate.bib}
\end{document}